\documentclass{elsart}
\usepackage{epsf}
\usepackage{amsmath}
\usepackage{graphicx}
\usepackage{amsfonts}
\usepackage{amssymb}

\begin{document}
\begin{frontmatter}
\title{Thermodynamic potential of the Periodic Anderson Model with the X-boson method: Chain Approximation}
\author[UFF]{R. Franco \thanksref{CNPq}}
\author[UFF]{M. S. Figueira \thanksref{CNPq}}
\author[UNICAMP]{M. E. Foglio \thanksref{CNPq} \thanksref{FAPESP}}
\address[UFF]{Instituto de F\'{\i}sica, Universidade Federal Fluminense (UFF),
Av. Litor\^{a}nea s/n, CEP 24210-340, Niter\'oi, RJ, Brazil}
\address[UNICAMP]{Instituto de F\'{\i}sica, Universidade Estadual de
Campinas,\\ 13083-970 Campinas, S\~{a}o Paulo, Brasil}
\thanks[CNPq] {Partially supported by the National Research Council (CNPq)}
\thanks[FAPESP] {Partially supported by the S\~{a}o Paulo State
Research Foundation (FAPESP)}
\thanks
[EMAIL] {Email adresses: rfranco@if.uff.br, figueira@if.uff.br, foglio@ifi.unicamp.br}
\begin{abstract}
The Periodic Anderson Model (PAM) in the $U\rightarrow\infty$
limit has been studied in a previous work employing the cumulant
expansion with the hybridization as perturbation (M. S. Figueira,
M. E. Foglio and G. G. Martinez, Phys. Rev. B \textbf{50}, 17933
(1994)). When the total number of electrons $N_{t}$ is calculated
as a function of the chemical potential $\mu$ in the  ``Chain
Approximation'' (CHA), there are three values of the chemical
potential $\mu$ for each $N_{t}$ in a small interval of $N_{t}$ at
low $T$ (M. S Figueira, M. E Foglio, Physica A 208 (1994)). We
have recently introduced the ``X-boson'' method, inspired in the
slave boson technique of Coleman, that solves the problem of non
conservation of probability (completeness) in the CHA as well as
removing the spurious phase transitions that appear with the slave
boson method in the mean field approximation. In  the present
paper we show that the X-boson method solves also the problem of
the multiple roots of $N_{t}(\mu)$ that appear in the CHA.
\end{abstract}
\end{frontmatter}

\newpage

\section{Introduction\bigskip}

In the Periodic Anderson Model (PAM) there are four local states at each site
$j$ of the lattice (identified by a single index in this work): the vacuum
state \ $\left|  \ j,0\right\rangle $, the two states $\left|  \ j,\sigma
\right\rangle $ of one electron with spin component $\sigma=\pm1$ and the
state $\left|  \ j,2\right\rangle $ with two local electrons \cite{HewsonB}.
\ In the limit of infinite Coulomb repulsion $U\rightarrow\infty$, \ the state
$\left|  \ j,2\right\rangle $ is empty, and we shall use the Hubbard operators
\cite{Hubbard4} to project it out from the space of local states at site $j$.
The Hamiltonian of the system is then \cite{FFM}:%

\begin{align}
H  &  =\sum_{\mathbf{k},\sigma}E_{\mathbf{k},\sigma}c_{\mathbf{k},\sigma
}^{\dagger}c_{\mathbf{k},\sigma}+\sum_{j,\sigma}\ \varepsilon_{f,j\sigma
}X_{j,\sigma\sigma}+\nonumber\\
&  \sum_{j,\sigma,\mathbf{k}}\left(  V_{j,\sigma,\mathbf{k}}X_{j,0\sigma
}^{\dagger}c_{\mathbf{k},\sigma}+V_{j,\sigma,\mathbf{k}}^{\ast}c_{\mathbf{k}%
,\sigma}^{\dagger}X_{j,0\sigma}\right)  , \label{Eq.3}%
\end{align}

\noindent where the first term is the Hamiltonian of the conduction electrons
($c$-electrons), the second term describes independent localized electrons
($f$-electrons), and the last term is the hybridization Hamiltonian giving the
interaction between the $c$-electrons and the $f$-electrons.

The $X$ Hubbard operators do not satisfy the usual commutation relations so
that diagrammatic methods based on Wick's theorem are not applicable, and one
has to use product rules instead:%
\begin{equation}
X_{j,ab}.X_{j,cd}=\delta_{b,c}X_{j,ad}. \label{HO}%
\end{equation}
The identity decomposition in the reduced space of local states at site $j$ is
then%
\begin{equation}
X_{j,00}+X_{j,\sigma\sigma}+X_{j,\overline{\sigma}\overline{\sigma}}=I_{j},
\label{Eq.1}%
\end{equation}

\noindent where $\overline{\sigma}=-\sigma$, and the three $X_{j,aa}$ are the
projectors into $\mid j,a\rangle$. Because of the translational invariance,
the occupation numbers $n_{j,a}=<X_{j,aa}>$ \ satisfy $n_{j,a}$=$n_{a}$
(independent of j), and from Eq. (\ref{Eq.1}) we obtain the ``completeness''
relation
\begin{equation}
n_{0}+n_{\sigma}+n_{\overline{\sigma}}=1. \label{Eq.2}%
\end{equation}

The occupation numbers can be calculated from appropriate Green's functions
(GF), and it has been found that Eq. (\ref{Eq.2}) is not usually satisfied
when the $n_{a}$ are calculated with approximate cumulant Green's functions
(GF) \cite{Ufinito}. An approximation with this behavior is the ``Chain
Approximation'' (CHA), which was first obtained by Hewson
\cite{Hewson,Enrique}. This approximation has interesting properties: it is
$\Phi-$derivable \cite{Baym61,ChainPhi}, and it is also the most general
cumulant expansion with only second order cumulants. We have employed several
procedures to restore completeness to the CHA: renormalization of the
one-electron Green's functions (GF) or adding diagrams to the GF in the CHA.
The second method lead us to a conjecture on a systematic way of achieving
completeness by adding a set of diagrams to an arbitrary family
\cite{ChainPhi}. An alternative to these techniques was inspired in the mean
field treatment of the slave boson technique \cite{Coleman84,Read87}, in which
the correlated problem is transformed into an uncorrelated one with one
condition that forces to zero the occupation of $\left|  \ j,2\right\rangle $
states. This condition turns out to be just our Eq. (\ref{Eq.2}), and
following Coleman we minimized the free energy of the system calculated with
the CHA but forcing the validity of Eq. (\ref{Eq.1}), that implies the
completeness. This is the essence of the X-boson method
\cite{JMMM2001,X-boson} and, differently from the slave boson treatment, the
correlation is kept at the final stage because it is intrinsic to the CHA
\cite{FoglioSUN}. An important consequence of this fact is that the spurious
phase transition that appears in the slave boson treatment for several regions
of the system parameters (for intermediate temperatures or when $\mu>>E_{f}$)
\cite{Coleman87}, disappears completely from our treatment. The results of the
X-boson method are fairly close to those obtained by the slave boson treatment
in the region of its validity, while the X-boson method gives results that are
close to those of the CHA when the slave boson is not valid any more (at high
temperatures and when $\mu>>\varepsilon_{f}$ ).

A rather inconvenient aspect of the CHA is that its results show a region of
instability \cite{Physica A 94}, apparent because the dependence of $N_{t}$
with $\mu$ shows multiple values of $\mu$ for a given $N_{t}$ within a small
interval of values of $N_{t}$, and this leads to negative compressibility. The
main result of the present work is to show that this difficulty of the CHA is
removed by the X-boson treatment, and $N_{t}(\mu)$ becomes then a monotonous function.

The paper is organized as follows: the X-boson approach is presented in
section \ref{X-boson Method}, the GF in the CHA are given in section \ref{CHA
GFs}, and in section \ref{Free Energy} we describe the calculation of the free
energy. Our results and conclusions are presented in section \ref{results}.

\section{X-boson Cumulant Method}

\label{X-boson Method}

In Coleman's ``slave boson'' method \cite{Coleman84,Read87}, the Hubbard X
operators are written as a product of ordinary bosons and
fermions:$\ \ X_{j,oo}\rightarrow b_{j}^{+}b_{j}$, $X_{j,o\sigma}\rightarrow
b_{j}^{+}f_{j,\sigma}$, $X_{j,\sigma o}\rightarrow f_{j,\sigma}^{+}b_{j}$, and
a condition, that is equivalent to Eq. (\ref{Eq.2}), is imposed to avoid
states with two electrons at each site $j$. In the spirit of the mean field
approximation $b_{i}^{+}\rightarrow<b_{i}^{+}>=r$ and the method of Lagrangian
multipliers is employed to minimize the free energy subject to that condition.
The problem is then reduced to an uncorrelated Anderson lattice with
renormalized hybridization $V\rightarrow rV$ and $f$ level $\varepsilon
_{f}\rightarrow\varepsilon_{f}+\lambda$, and the conservation of probability
in the space of local states is automatically satisfied because they are
described by Fermi operators.

The approximate GF obtained by the cumulant expansion \cite{FFM,Infinite}
do not usually conserve
probability (i.e. they do not satisfy Eq. (\ref{Eq.2})), and the procedure we
adopt to recover this property in the X-boson method is to introduce
\begin{equation}
R\equiv<X_{j,oo}>=<b_{j}^{+}b_{j}>, \label{Eq.5}%
\end{equation}
as variational parameter, and to modify the approximate GF so that it
minimizes an adequate thermodynamic potential while being forced to satisfy
Eq. (\ref{Eq.2}). To this purpose we add to Eq. (\ref{Eq.3}) the product of
each Eq. (\ref{Eq.2}) into a Lagrange multiplier $\Lambda_{j}$, and employ
this new Hamiltonian to generate the functional that shall be minimized by
employing Lagrange's method. To simplify the calculations we use a constant
hybridization $V$, as well as site independent local energies ${\varepsilon
}_{f,j,\sigma}={\varepsilon}_{f,\sigma}$ and Lagrange parameters $\Lambda
_{j}=\Lambda$. We then obtain a new Hamiltonian with the same form of Eq.
(\ref{Eq.3}):
\begin{align}
H  &  =\sum_{{\vec{k}},\sigma}\ E_{{\vec{k}},\sigma}\ c_{{\vec{k}},\sigma
}^{\dagger}c_{{\vec{k}},\sigma}+\nonumber\\
&  \sum_{j,\sigma}\tilde{\varepsilon}_{f,\sigma}X_{j,\sigma\sigma}%
+N_{s}\Lambda(R-1)+\nonumber\\
&  V\sum_{j,{\vec{k}},\sigma}\left(  X_{j,0\sigma}^{\dagger}\ c_{{\vec{k}%
},\sigma}+c_{{\vec{k}},\sigma}^{\dagger}\ X_{j,0\sigma}\right)  , \label{Eq.7}%
\end{align}
\noindent but with renormalized localized energies%
\begin{equation}
\tilde{\varepsilon}_{f,\sigma}=\varepsilon_{f,\sigma}+\Lambda.
\end{equation}
The parameter%
\begin{equation}
R=1-\sum_{\sigma}<X_{\sigma\sigma}> \label{Eq. 9}%
\end{equation}
is now varied independently to minimize the thermodynamic potential, choosing
$\Lambda$ so that Eq. (\ref{Eq.2}) be satisfied. While at this stage the
electrons in the slave boson Hamiltonian have lost all the correlations, the
Eq. (\ref{Eq.7}) is still in the projected space and it is not necessary to
force the correlations with an extra condition. On the other hand we do not
have an exact solution for this new problem, and we then consider the most
simple approximation obtained within the cumulant formalism, the Chain
approximation (CHA) \cite{Hewson,Enrique}. The need of minimizing a
thermodynamic potential arises because the completeness relation is not
automatically satisfied for approximate cumulant solutions, and although the
two procedures are formally very similar, they have a rather different meaning.

The Grand Canonical Ensemble of electrons is employed in the present
treatment, and instead of Eq.~(\ref{Eq.3}) we have to use
\begin{equation}
\mathcal{H}=H-\mu\left\{  \sum_{\mathbf{k,\sigma}}c_{\mathbf{k,\sigma}%
}^{\dagger}c_{\mathbf{k,\sigma}}+\sum_{ja}\nu_{a}X_{j,aa}\right\}  ,
\label{E2.2}%
\end{equation}

\noindent where $\nu_{a}=0,1$ is the number of electrons in state $\mid a>$.
It is then convenient to define
\begin{equation}
\varepsilon_{j,a}=E_{f,j,a}-\mu\nu_{a} \label{E2.3a}%
\end{equation}
and
\begin{equation}
\varepsilon_{\mathbf{k\sigma}}=E_{\mathbf{k\sigma}}-\mu\ , \label{E2.3b}%
\end{equation}
because $E_{f,j,a}$ and $E_{\mathbf{k,\sigma}}$ appear only in that form in
all the calculations. The exact and unperturbed averages of the operator $A$
are denoted in what follows by $<A>_{\mathcal{H}}$ and $<A>$ respectively.

\section{The Chain Approximation Green's Functions}

\label{CHA GFs}

The GF in the Chain approximation (CHA) are given by \cite{ChainPhi,X-boson}:%
\begin{equation}
G_{\mathbf{k}\sigma}^{ff}(z_{n})=\frac{-D_{\sigma}\left(  z_{n}-\varepsilon
_{\mathbf{k}\sigma}\right)  }{\left(  z_{n}-\varepsilon_{f,\sigma}\right)
\left(  z_{n}-\varepsilon_{\mathbf{k}\sigma}\right)  -|V_{\sigma}%
(\mathbf{k})|^{2}D_{\sigma}}, \label{Eqn.11}%
\end{equation}%
\begin{equation}
G_{\mathbf{k}\sigma}^{cc}(z_{n})=\frac{-\left(  z_{n}-\varepsilon_{f,\sigma
}\right)  }{\left(  z_{n}-\varepsilon_{f,\sigma}\right)  \left(
z_{n}-\varepsilon_{\mathbf{k}\sigma}\right)  -|V_{\sigma}(\mathbf{k}%
)|^{2}D_{\sigma}}, \label{Eqn.12}%
\end{equation}%
\begin{equation}
G_{\mathbf{k}\sigma}^{fc}(z_{n})=\frac{-\ D_{\sigma}V_{\sigma}(\mathbf{k}%
)}{\left(  z_{n}-\varepsilon_{f,\sigma}\right)  \left(  z_{n}-\varepsilon
_{\mathbf{k}\sigma}\right)  -|V_{\sigma}(\mathbf{k})|^{2}D_{\sigma}},
\label{Eqn.13}%
\end{equation}
where%

\begin{equation}
D_{\sigma{^{\prime}}}=<X_{oo}>+<X_{\sigma{^{\prime}}\sigma{^{\prime}}%
}>=R+n_{f\sigma{^{\prime}}}.
\end{equation}
and $\sigma{^{\prime}}=\sigma,\overline{\sigma}$. The $\hspace{0.1cm}$ lattice
$\hspace{0.1cm}$ slave-boson $\hspace{0.1cm}$ GF are recovered$\hspace{0.1cm}$
if $\hspace{0.1cm}$ we put $\hspace{0.1cm}$ $D_{\sigma}=1$ and $V_{\mathbf{k}%
}\rightarrow rV_{\mathbf{k}}=\overline{V}_{\mathbf{k}}$ in Eqs.(12,13,14). The
usual GF for the lattice in the CHA are obtained if we use the bare
$D_{\sigma}$, while $D_{\sigma}$ must be calculated self-consistently in the
X-boson approach.

In an earlier paper \cite{Physica A 94} we considered the atomic limit of the
PAM, i.e. with a conduction band of zero width, to study both the free energy
and $N_{t}$ as a function of $\mu$, but here we shall consider the PAM with a
wide band.

\section{Calculation of the Lattice Helmholtz Free Energy}

\label{Free Energy}

When the total number of electrons $N_{t}$, the temperature $T$ and the volume
$V_{s}$ are kept constant the equilibrium state corresponds to a minimum of
the Helmholtz free energy, but the same state of equilibrium is obtained by
minimizing the thermodynamic potential $\Omega=-k_{B}T\ln(Q)$, (where $Q$ is
the Grand Partition Function) at constant $T$, $V_{s}$, and chemical potential
$\mu$ (this result is easily obtained by employing standard thermodynamic
techniques). We shall then employ $\Omega$ as the thermodynamic potential that
is minimized in the X-boson method with Eq. (\ref{Eq.2}) as constraint. The
Helmholtz free energy $F$ is then given by%
\begin{equation}
F=N_{t}\ \mu+\Omega, \label{Eq.14.0}%
\end{equation}
and our first step would be the calculation of $\Omega$.

A convenient way to obtain $\Omega$ is to employ the method of $\xi$ parameter
integration \cite{Physica A 94,Abrikosov,Doniach}. This method introduces a
$\xi$ dependent Hamiltonian $H({\xi})=H_{o}+{\xi}H_{1}$ through a coupling
constant ${\xi}$ (with $0\leq{\xi}\leq1$), where $H_{1}$ is the hybridization
in our case. One obtains \cite{X-boson}%
\begin{equation}
\Omega=\Omega_{o}+\int_{0}^{1}d{\xi}<H_{1}(\xi)>_{\xi}, \label{Eqn.15}%
\end{equation}
\noindent where $<A>_{{\xi}}$ is the ensemble average of an operator $A$ for a
system with Hamiltonian $H({\xi})$ and the given values of $\mu,$ $T$, and
$V_{s}$, while $\Omega_{o}$ is the thermodynamic potential of the system with
${\xi}=0$. This value of ${\xi}$ corresponds to a system without
hybridization, and one obtains (in the absence of magnetic field
$\varepsilon_{\mathbf{k\sigma}}=\varepsilon_{\mathbf{k}}$ and $\widetilde
{\varepsilon}_{f\sigma}=\widetilde{\varepsilon}_{f}$)
\begin{align}
\Omega_{o}  &  =-\frac{2}{\beta}{\sum_{\mathbf{k}}}\ln\left[  1+\exp
(-\beta\varepsilon_{\mathbf{k}})\right] \nonumber\\
&  -\frac{N_{s}}{\beta}\ln\left[  1+2\exp(-\beta\widetilde{\varepsilon}%
_{f})\right]  +N_{s}\Lambda(R-1), \label{Eqn.16}%
\end{align}
and to calculate $\Omega$ in Eq.(\ref{Eqn.15}) we use
\begin{equation}
\left\langle H_{1}\right\rangle _{\xi}=2Re\left[  \sum_{\mathbf{k}\sigma
}V_{j,\mathbf{k},\sigma}^{\ast}\left\langle c_{\mathbf{k}\sigma}^{\dagger
}X_{0\sigma}\right\rangle _{\xi}\right]  . \label{Eqn.17}%
\end{equation}
Employing standard Green's functions techniques \cite{X-boson} we find%
\begin{equation}
\left\langle H_{1}\right\rangle _{\xi}=\frac{1}{\pi}\int\limits_{-\infty
}^{\infty}d\omega\ \ n_{F}(\omega)\sum_{\mathbf{k,}\sigma}Im{\frac
{\xi|V_{\sigma}(\mathbf{k})|^{2}D_{\sigma}}{\left(  \omega^{+}-\varepsilon
_{f\sigma}\right)  \left(  \omega^{+}-\varepsilon_{\mathbf{k}\sigma}\right)
-{\xi}^{2}|V_{\sigma}(\mathbf{k})|^{2}D_{\sigma}},} \label{Eqn.19}%
\end{equation}

where $n_{F}(x)=1/\left[  1+\exp(\beta x)\right]  $ is the Fermi-Dirac
distribution and $\omega^{+}=\omega+i0$. This equation has an interesting
scaling property: it is equal to the corresponding expression of the
uncorrelated system for the scaled parameter $\overline{V}_{j,\mathbf{k}%
,\sigma}=\sqrt{D_{0\sigma}}V_{j,\mathbf{k},\sigma}$ (it is enough to remember
that by replacing $D_{0\sigma}=1$ in the GF of the CHA one obtains the
corresponding GF of the uncorrelated system). Rather than performing the
${\xi}$ and $\omega$\ integrations, we shall use the value of the $\Omega^{u}$
for the uncorrelated system with $\overline{V}_{j,\mathbf{k},\sigma}%
=\sqrt{D_{0\sigma}}V_{j,\mathbf{k},\sigma}$ and employ Eq. (\ref{Eqn.15}) to
calculate $\int_{0}^{1}d{\xi}<H_{1}^{u}({\xi})>_{{\xi}}=\Omega^{u}-\Omega
_{o}^{u}$, where
\begin{equation}
\Omega_{o}^{u}=\frac{-2}{\beta}\left[  {\sum_{\mathbf{k}}}\ln\left[
1+\exp(-\beta\varepsilon_{\mathbf{k}})\right]  +N_{s}\ln\left[  1+\exp
(-\beta\widetilde{\varepsilon}_{f})\right]  \right]  +N_{s}\Lambda(R-1)
\label{Eqn.19a}%
\end{equation}
is the $\Omega^{u}$ for $\overline{V}_{\mathbf{k},\sigma}=0$. The uncorrelated
Hamiltonian for the lattice problem is then%
\begin{align}
H^{u}  &  =\sum_{\mathbf{k},\sigma}\varepsilon_{\mathbf{k},\sigma
}c_{\mathbf{k},\sigma}^{\dagger}c_{\mathbf{k},\sigma}+\sum_{j,\sigma
}\widetilde{\varepsilon}_{f}\ f_{{j},\sigma}^{\dagger}f_{{j},\sigma}%
+N_{s}\Lambda(R-1)\nonumber\\
&  +\sum_{j,\mathbf{k},\sigma}\left(  \overline{V}_{\mathbf{k},\sigma
}f_{j,\sigma}^{\dagger}c_{\mathbf{k},\sigma}+\overline{V}_{j,\mathbf{k}%
,\sigma}^{\ast}c_{\mathbf{k},\sigma}^{\dagger}f_{j,\sigma}\right)  ,
\label{Eqn.20}%
\end{align}
and this Hamiltonian can be easily diagonalized. The corresponding
${\mathcal{H}}^{u}$ can be written
\begin{equation}
{\mathcal{H}}^{u}=\sum_{i,\sigma}\omega_{i,\sigma}\ \alpha_{i,\sigma}%
^{\dagger}\alpha_{i,\sigma}\ +\Lambda(R-1), \label{Eqn.21}%
\end{equation}
where $\alpha_{i,\sigma}^{\dagger}$ ($\alpha_{i,\sigma}$) are the creation
(destruction) operators of the composite particles of energies $\omega
_{i,\sigma}$ (there are $N_{s}+1$ states for each spin $\sigma$). The
calculation of $\Omega^{u}$ is then straightforward \cite{X-boson}:
\begin{equation}
\Omega=\overline{\Omega}_{0}+\frac{-1}{\beta}\sum_{i,\sigma}\ln\left[
1+\exp(-\beta\ \omega_{i,\sigma})\right]  +\Lambda(R-1), \label{Eqn.23}%
\end{equation}
where
\begin{equation}
\overline{\Omega}_{0}\equiv\Omega_{o}-\Omega_{o}^{u}=-\frac{N_{s}}{\beta}%
\ln\left[  \frac{1+2\exp(-\beta\tilde{\varepsilon}_{f})}{\left(  1+\exp
(-\beta\tilde{\varepsilon}_{f})\right)  ^{2}}\right]  , \label{Eqn.24}%
\end{equation}
and the eigenvalues $\omega_{i,\sigma}$ of the ${\mathcal{H}}^{u}$ are just
given by the poles of the GF in the CHA (Eq.(\ref{Eqn.11})). In the present
case the $\omega_{i,\sigma}$ can be calculated analytically, because the
Hamiltonian for each spin component $\sigma$ is reduced into $N_{s}$ matrices
$2\times2$, and one finds%

\begin{equation}
\omega_{i,\sigma}=\omega_{\pm,\sigma}(\mathbf{k})=\frac{1}{2}\left(
\varepsilon_{\mathbf{k},\sigma}+\widetilde{\varepsilon}_{f}\right)  \pm
\frac{1}{2}\sqrt{\left(  \varepsilon_{\mathbf{k},\sigma}-\widetilde
{\varepsilon}_{f}\right)  ^{2}+4\left|  V_{\sigma}(\mathbf{k})\right|
^{2}D_{\sigma}}. \label{Eqn.32}%
\end{equation}

The parameter $\Lambda$ is obtained minimizing ${\Omega}$ with respect to $R$
\cite{X-boson}. To simplify the calculations we shall consider a conduction
band with a constant density of states, width $W=2D$, an hybridization
constant $V_{\sigma}(\mathbf{k})=V$, and $\varepsilon_{\mathbf{k},\sigma
}=\varepsilon_{\mathbf{k}}$; we then obtain%

\begin{equation}
\Lambda=\frac{V^{2}}{D}\int_{-D}^{D}d\varepsilon_{\mathbf{k}}\frac
{n_{F}\left(  \omega_{+}(\mathbf{k})\right)  -n_{F}\left(  \omega
_{-}(\mathbf{k})\right)  }{\sqrt{\left(  \varepsilon_{\mathbf{k}}%
-\tilde{\varepsilon}_{f}\right)  ^{2}+4V^{2}D_{\sigma}}}.
\end{equation}

All the
correlation effects in \ Eq. (\ref{Eqn.23}) are included in $\Omega_{o}$ (cf.
Eq. (16)), and the hybridization redistributes the quasi-particle energies in
the same way that an uncorrelated system with $\overline{V}_{j,\mathbf{k}%
,\sigma}=\sqrt{D_{0\sigma}}V_{j,\mathbf{k},\sigma}$ would do; one then expects
Fermi liquid behavior of the quasi-particles \cite{X-boson}.

\section{Results and Conclusions}

\label{results}

Employing the CHA to calculate the GF of the PAM one discovers an undesirable
feature in several regions of the parameter space and for sufficiently low $T$
\cite{Physica A 94}. When the total number of particles in the system $N_{t}$
is plotted against the chemical potential $\mu$ one finds that there are three
possible values of $\mu$\ for each $N_{t}$ in an interval of $N_{t}$,
\ originating an instability in a small interval of $\mu$, characterized by a
negative derivative of $N_{t}$ with respect of $\mu$. The problem was first
observed in a wide conduction band, but further analysis was done for a band
of zero width \cite{Physica A 94}, because one could then obtain an exact
solution of the problem and compare with the approximate result. The value of
the free energy $F$ was then calculated for the three states to analyze their
relative stability. Here we shall consider the problem for a wide conduction
band, and compare the results of the X-boson with those obtained with the CHA
and with the slave-boson. The main result of the present work is that the
X-boson does not present the multiple solutions of the CHA.

\begin{figure}[th]
\centering
\epsfysize=2.7truein \centerline{\epsfbox{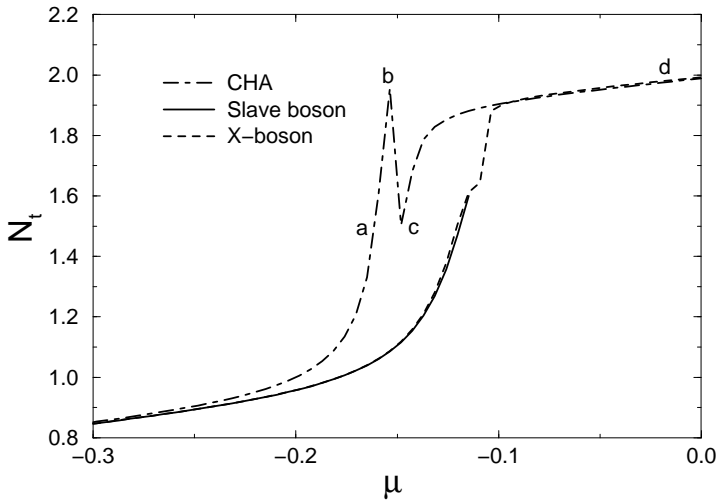}}\caption{Total Number
of particles $N_{t}$ vs. $\mu$,\hspace{0.1cm} in the three approaches with the
following parameters: $\varepsilon_{f}=-0.15$; $W=\pi$; $V=0.1$;
$T=0.001$. The slave boson breaks down when $\mu \approx -0.12.$}%
\label{figrede1}%
\end{figure}

In Figure~\ref{figrede1} we plot $N_{t}(\mu)$ for CHA, slave boson
and X-boson for the same parameters used in \cite{Physica A 94}.
The multiple solutions of the CHA are apparent, as well as the
thermodynamically unstable region (part b-c of the
Fig.~\ref{figrede1}, where $N_{t}$ decreases when $\mu$
increases), i.e. the same behavior that was observed for the
atomic limit of the PAM \cite{Physica A 94}. The problem of
thermodynamic unstable solutions does not appear in the slave
boson and X-boson, but the slave boson breaks down at higher $\mu$
($\mu>>\tilde{\varepsilon_{f}}$,\hspace{0.1cm} magnetic regime)
while the X-boson presents a continuous evolution in all regimes
of the model.

\begin{figure}[th]
\centering \epsfysize=2.7truein \centerline
{\epsfbox{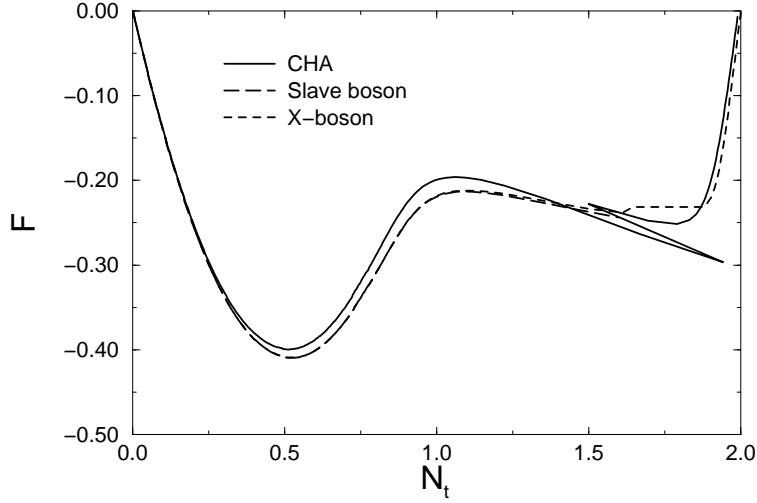}} \caption{Helmholtz free energy $F$ vs.
$N_{t}$,\hspace{0.1cm} for the X-boson, Slave boson and CHA
approaches with the same parameters of Figure \ref{figrede1}.}
\label{figrede2}
\end{figure}

In Figure \ref{figrede2} we present the Helmholtz free energy $F$ vs. $N_{t}$
in the slave boson, X-boson, and CHA approximations. The CHA present three
values of free energy $F$ for the same $N_{t}$ that have three values of $\mu
$, while this problem is absent in the X-boson and slave boson methods.

\begin{figure}[th]
\centering
\epsfysize=2.7truein \centerline{\epsfbox{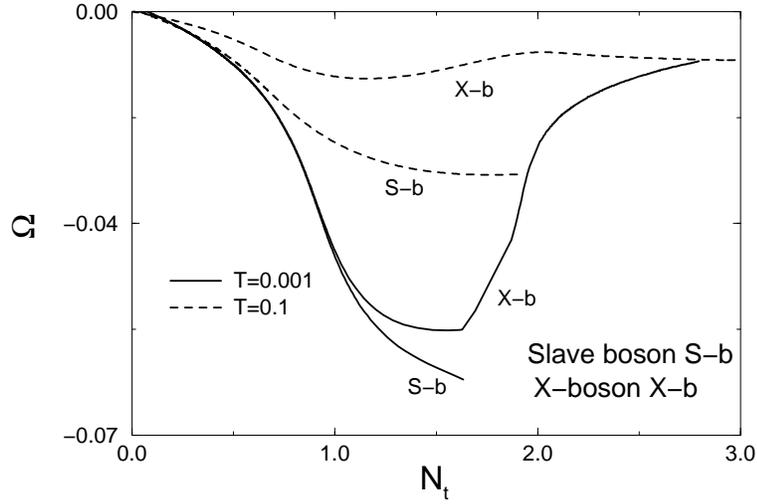}}\caption{Grand
thermodynamic potential $\Omega$ vs. $N_{t}$,\hspace{0.1cm} for the slave
boson and X-boson approaches with the same parameters of Figure \ref{figrede1},
but for two values of the temperature:  $T=0.1$ and $T=0.001$.}%
\label{figrede3}%
\end{figure}

In Figure \ref{figrede3} we plot the grand thermodynamic potential $\Omega$
vs. $N_{t}$ for the slave boson and X-boson,$\hspace{0.1cm}$ at two values of
$T$. $\hspace{0.1cm}$ In this plot the grand thermodynamic potential $\Omega$
for the slave boson is always lower or equal than for the X-boson, and its
values increase with temperature in the two methods; its minimum values are in
the Kondo region at the lowest $T.$ The values of $\Omega$\ are different at
the larger value of $T$ for all $N_{t}$.

\begin{figure}[th]
\centering
\epsfysize=2.7truein \centerline{\epsfbox{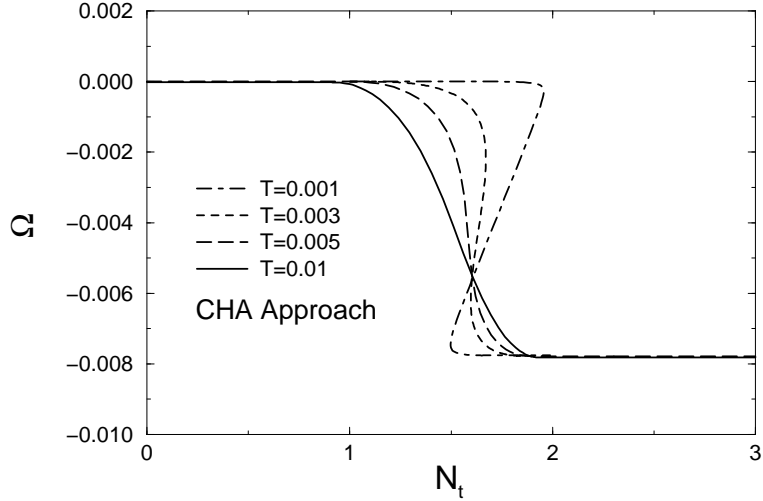}}\caption{Grand
thermodynamic potential $\Omega$ vs. $N_{t}$,\hspace{0.1cm} for the CHA
approach with the same parameters of Figure \ref{figrede1},
but for several values of the temperature.}%
\label{figrede4}%
\end{figure}

In Figure \ref{figrede4} we show that $\Omega(N_{t})$ has multiple values for
the CHA at $T=0.001$ and $T=0.003$ when $N_{t}\backsim1.5$, while that
behavior is absent at higher $T$. Note that there is no minimum of $\Omega$ in
the CHA, and that this quantity changes with $N_{t}$ between maximum and
minimum values that are independent of $T$.

\begin{figure}[th]
\centering \epsfysize=2.7truein
\centerline{\epsfbox{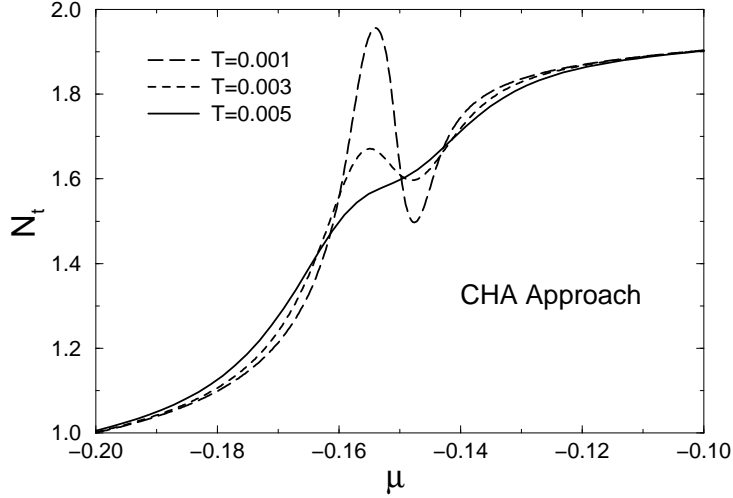}} \caption{Total Number of
particles $N_{t}$ vs. $\mu$,\hspace{0.1cm} for the CHA approach
with the same parameters of Figure \ref{figrede1}
but at different temperatures values.}%
\label{figrede5}%
\end{figure}

In Figure \ref{figrede5} we plot $N_{t}$ vs. $\mu$ for the CHA at
different values of $T$, and the multiple values of $\mu$ for a
given $N_{t}$ disappear at the larger value of $T$.

\begin{figure}[th]
\centering
\epsfysize=2.7truein \centerline{\epsfbox{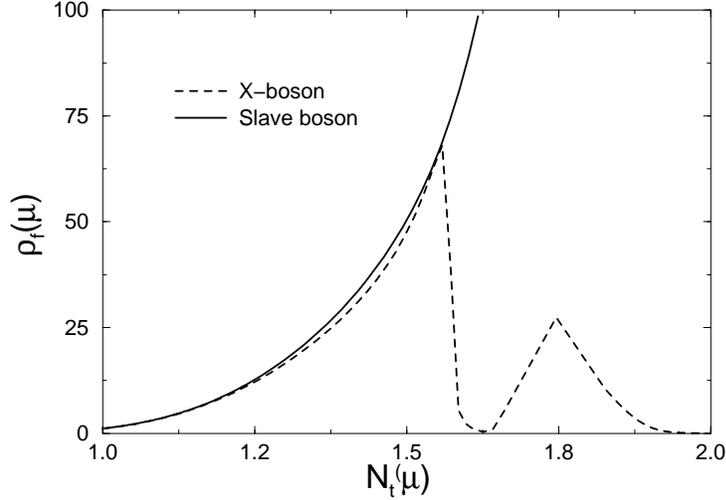}} \caption{Density of
states $f$ on the Fermi level $\mu$, $\rho_{f}(\mu)$ vs. $N_{t}(\mu)$%
,\hspace{0.1cm} for the X-boson and Slave boson approaches with
the same parameters of Figure \ref{figrede1}.}%
\label{figrede6}%
\end{figure}

In Figure \ref{figrede6} we plot $\rho_{f}(\mu)$ (the density of states $f$ on
the Fermi level $\mu)$ vs. $N_{t}$ for the X-boson and slave boson at
$T=0.001$, the lower value of temperature that we consider in Figure
\ref{figrede3}. From this figure is clear that the Kondo effect is present
when $N_{t}\approx1.6$ where $\rho_{f}(\mu)$ has a maximum in the two approaches.

\subsubsection{Conclusions}

Some years ago we studied the thermodynamic properties of the CHA
\cite{Physica A 94}, an approximation that was obtained by employing a
diagrammatic cumulant expansion for the PAM. We then found that for a region
of the system parameters at very low temperatures, there are three states with
different chemical potential $\mu$ for the same total number of electrons
$N_{t}$. We observed that one of these states is thermodynamically unstable
(part b-c of the Fig. \ref{figrede1}) because $N_{t}$ decreases when $\mu$
increases. The X-boson is essentially a generalized CHA approximation which
satisfies completeness through a minimization of the thermodynamic potential
as a function of the average occupation of the empty state while keeping Eq.
(\ref{Eq.1}) as a constraint. As it is shown in Figure (\ref{figrede1}) this
approach does not present thermodynamically unstable states or multiple
solutions for a given $N_{t}$. We conclude that the X-boson combines the
simplicity and usefulness of the CHA and slave boson methods, without their
more obvious defects: the presence of spurious phase transitions of the slave
boson and the absence of Kondo resonance, of multiple solutions and
instabilities, as well as the failure to satisfy completeness of the CHA.


\begin{thebibliography}{99}
\bibitem{HewsonB}{A. C. Hewson, The Kondo Problem to Heavy Fermions -
Cambridge Studies in Magnetism - Cambridge University Press, (1993).}

\bibitem{Hubbard4}J. Hubbard, Proc. R. Soc. London, Ser. A 285 (1965) 542; A
296 (1966) 82.

\bibitem{FFM}M. S. Figueira, M. E. Foglio and G. G. Martinez, Phys. Rev. B 50
(1994) 17933.

\bibitem{Ufinito}M. E. Foglio, M. S. Figueira, International Journal of Modern
Physics B 12 (1998) 837.

\bibitem{Hewson}A. C. Hewson J. Phys. C: Solid State Phys. 10 (1977) 4973.

\bibitem{Enrique}E. V. Anda, J. Phys. C: Solid State Phys. 14 (1981) L1037.

\bibitem{Baym61}G. Baym and L. P. Kadanoff Phys. Rev. 124 (1961) 287.

\bibitem{ChainPhi}M. S. Figueira and M. E. Foglio J. Phys.: Condens. Matter 8
(1996) 5017.

\bibitem{Coleman84}P. Coleman, Phys. Rev. B 29 (1984) 3035.

\bibitem{Read87}D. M Newns, N. Read Advances in Physics 36 (1987) 799.

\bibitem{JMMM2001}R. Franco, M. S. Figueira and M. E. Foglio J. Magn. Magn.
Mat 226-230 (2001) 194.

\bibitem{X-boson}R. Franco, M. S. Figueira and M. E. Foglio, unpublished. The
submited paper can be found in preprint (2001) [cond-mat/0109037].


\bibitem{FoglioSUN}M. E. Foglio, {\ Phys. Rev. B 43 (1991) 3192.}

\bibitem{Coleman87}P. Coleman, J. Magn. Magn. Mat 47\&48 (1985) 323.

\bibitem{Physica A 94}M. S. Figueira, M. E Foglio, Physica A 208 (1994) 279.

\bibitem{Infinite}  {\ M. E. Foglio, M. S. Figueira, J. Phys. A: Math. Gen.
30 (1997) 7879. }

\bibitem{Abrikosov}{\ A. A. Abrikosov, L. P. Gorkov and I. E. Dzyaloshinski,
Quantum Field Theoretical Methods in Statistical Physics (Pergamon, Oxford), (1965).}

\bibitem{Doniach}{\ S. Doniach and E. H. Sondheimer, Green's Functions for
Solid State Physicists (Benjamin, New York), (1974).}

\end{thebibliography}
\end{document}